\documentclass[journal]{IEEEtran}
\usepackage{cite}

\usepackage{graphicx}

\ifCLASSINFOpdf
\else
\fi
\usepackage{amsmath,amssymb}
\usepackage{algorithm}
\usepackage{algorithmic}
\begin{document}
%
\title{A Hardware-Oriented and Memory-Efficient Method for CTC Decoding}
%
%
%

\author{Siyuan~Lu,
        Jinming~Lu,
        Jun~Lin,~\IEEEmembership{Senior Member,~IEEE,}
        and~Zhongfeng~Wang,~\IEEEmembership{Fellow,~IEEE}
\thanks{The authors are with the School of Electronic Science and Engineering, Nanjing University,
Nanjing 210008, China (e-mail: sylu@smail.nju.edu.cn; jmlu@smail.nju.edu.cn; jlin@nju.edu.cn; zfwang@nju.edu.cn).}
}
\maketitle

\begin{abstract}
The Connectionist Temporal Classification (CTC) has achieved great success in sequence to sequence analysis tasks such as automatic speech recognition (ASR) and scene text recognition (STR). These applications can use the CTC objective function to train the recurrent neural networks (RNNs), and decode the outputs of RNNs during inference.
While hardware architectures for RNNs have been studied, hardware-based CTC-decoders are desired for high-speed CTC-based inference systems.
This paper, \textit{for~the~first~time}, provides a low-complexity and memory-efficient approach to build a CTC-decoder based on the beam search decoding. Firstly, we improve the beam search decoding algorithm to save the storage space. Secondly, we compress a dictionary (reduced from 26.02MB to 1.12MB) and use it as the language model. Meanwhile searching this dictionary is trivial. Finally, a fixed-point CTC-decoder for an English ASR and an STR task using the proposed method is implemented with C++ language. It is shown that the proposed method has little precision loss compared with its floating-point counterpart. Our experiments demonstrate the compression ratio of the storage required by the proposed beam search decoding algorithm are 29.49 (ASR) and 17.95 (STR).
\end{abstract}


\IEEEpeerreviewmaketitle
\section{Introduction}
\label{sec:introduction}
\IEEEPARstart{I}{n} most automatic speech recognition (ASR) tasks and some sequential tasks,
such as lipreading and scene text recognition, the lengths of output sequences are not fixed. Furthermore,
the alignment between input and output is unknown\cite{Graves2012Supervised}. To address this issue, Graves \textit{et al.}\cite{graves2006connectionist} provided the Connectionist Temporal Classification (CTC) objective function to infer this alignment automatically.
CTC is an output layer for recurrent neural networks (RNNs), which allows RNNs to be trained for sequence transcription tasks without requiring
a prior alignment between the input and target sequences\cite{Graves2014Towards}.

In ASR tasks, the traditional approach is based on HMMs\cite{rabiner1986an},
while recent works have shown great interest in building end-to-end models, using CTC-based deep RNNs. By training networks with large amounts of data, CTC-based models achieved great success\cite{Graves2014Towards},\cite{miao2015eesen},\cite{das2018advancing},
\cite{miao2016an},\cite{Zenkel2017Comparison}, \cite{salazar2019self}.
CTC is also widely used in other learning tasks such as handwriting recognition and scene text recognition, offering superior performance\cite{Graves2008Offline},\cite{Bluche2015Framewise},\cite{Shi2017AnET}.

In a learning task using CTC, models are always ended with a softmax layer where the element represents the probability of
emitting each label at a specific time step. After being trained with the CTC loss function, the output of the network needs a
CTC-decoder during inference. Since the probability of each label is temporally independent, a language model (LM)
can be integrated to improve the accuracy of CTC decoding.

\begin{figure}[ht]
    \centering
    \includegraphics[scale=0.66]{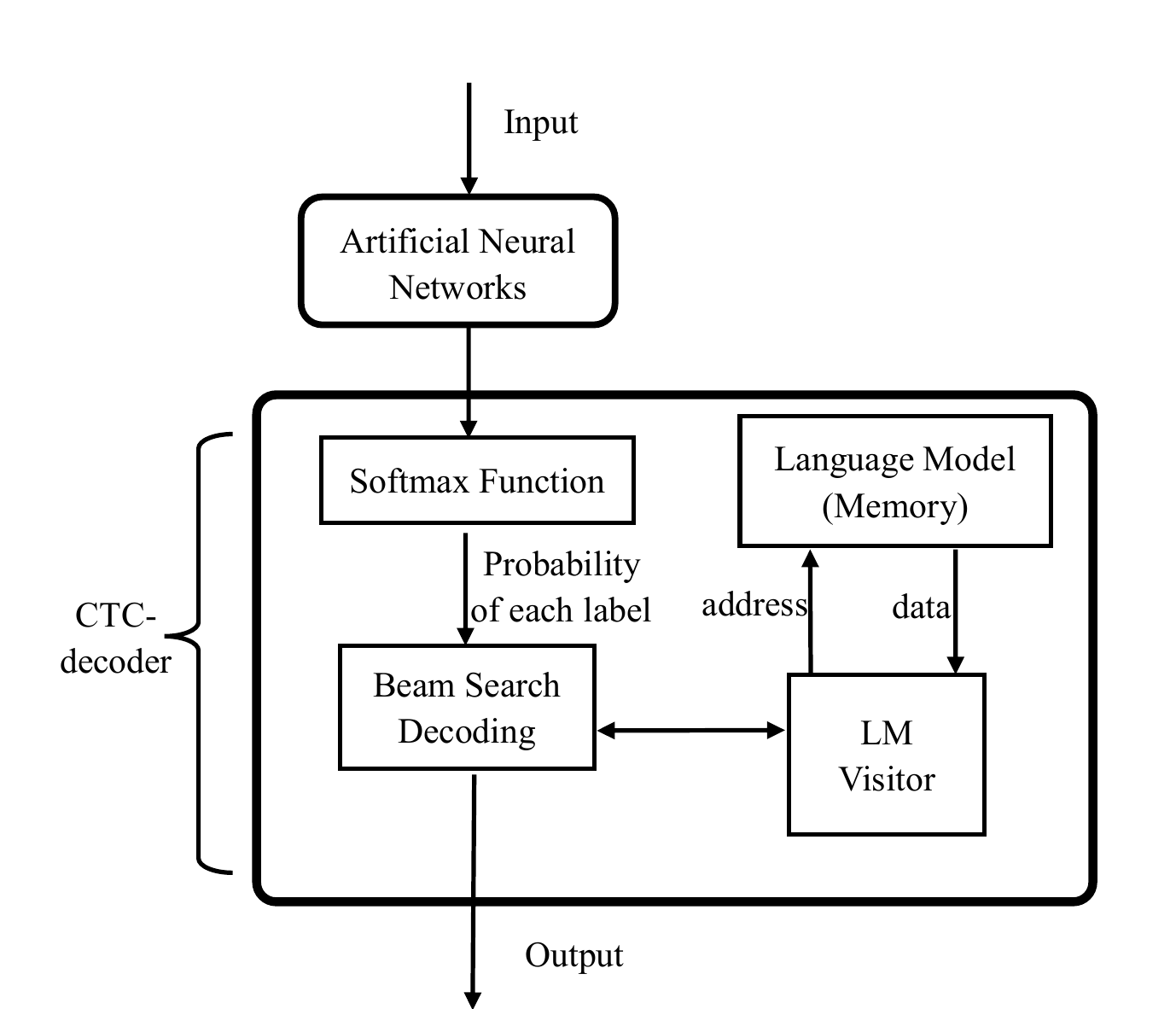}
    \newlength{\xfigwd}
    \caption{A sequence processing system using the CTC-decoder designed in this paper.}
    \label{fig:system framework}
\end{figure}

On one hand, compared with solutions based on CPUs and GPUs, hardware-based sequence to sequence systems can have lower power consumption and higher speed\cite{tabani2017ultra}\cite{yazdani2016ultra}\cite{price201714}.
On the other hand, CTC-decoder is an essential part of a system including CTC-trained neural networks.
The outputs of these neural networks cannot be combined into the target output sequences directly without a CTC-decoder.
Considering that recent works on hardware-based RNNs have made great progress\cite{Han2017ESE}\cite{Wang2017Accelerating}\cite{wang2018c},
hardware-based CTC-decoders are desired for high-speed CTC-based inference systems, which can make these systems more efficient.
In addition, the softmax function, which is also widely used in various neural networks\cite{Yuan2017Efficient},
involves expensive division and exponentiation units. So a low-complexity hardware architecture design of softmax is also in demand.

A sequence processing system using the CTC-decoder designed in this paper is shown in Fig. \ref{fig:system framework}.
The system consists of two concatenated stages: the neural network and the CTC-decoder, which can be run in pipline.
The network usually takes more cycles than the CTC-decoder to process a set of data\cite{Wang2017Accelerating},
so we do not need the decoder to run at high throughput. Thus, this decoder is designed to be serial to consume less computational resources.

There is no existing work on hardware-oriented algorithm nor hardware architecture for CTC decoding based on the beam search.
This paper, \textit{for~the~first~time},
provides a hardware-oriented CTC decoding approach,
employing the CTC beam search decoding with a dictionary as its LM.
Our contributions can be summarized as follows:

\begin{itemize}
  \item[1)] We improve the beam search decoding algorithm in \cite{Graves2014Towards}.
  We choose this decoding method as it can integrate all kinds of LMs.
  We reduce the memory size used in decoding as much as possible.
  The improvement is suitable for both software and hardware decoding, regardless of the kind of LM.
  We further point out that some components can be reused to reduce the hardware complexity.
  \item[2)] Several techniques are exploited to compress the size of a dictionary used as the LM in CTC decoding.
  By using these techniques, we compress the size of an English dictionary with 191,735 words from 26.05MB to 1.12MB.
  Meanwhile, we propose a low-complexity algorithm for the LM visitor.
  Our work on how to compress a dictionary is also useful when more complex LMs are used, as most of these LMs are based on a dictionary.
  \item[3)]We use C++ language to implement a fixed-point CTC decoder applying the hardware-friendly approach for softmax and the improved beam search decoding algorithm.
  In our experiments, the fixed-point decoder achieved nearly identical accuracy to the floating-point decoder in ASR and scene text recognition(STR) tasks,
  with the compression ratio of the storage are 29.49 and 17.95, respectively.

\end{itemize}

The RNN+CTC model is widely used, and the CTC beam search decoding algorithm is one of the most popular decoding methods\cite{Zenkel2017Comparison}. However, the original beam search algorithm consumes a lot of memory space, making us believe that reducing storage consumption is very necessary. The proposed CTC decoding method is useful in improving any CTC-based inference systems, no matter whether it is software-based or hardware-based. Although a complete hardware implementation for the proposed CTC-decoder has not been finished yet (which will be conducted in the future work), we have implemented quantized CTC-decoders in the experiments to prove this.

The rest of this paper is organized as follows. Section \uppercase\expandafter{\romannumeral2} gives a brief review of CTC,
the beam search algorithm, and the CTC beam search decoding algorithm.
Several algorithmic strength reduction strategies applied in designing a low-complexity architecture for softmax are also introduced in Section \uppercase\expandafter{\romannumeral2}.
Section \uppercase\expandafter{\romannumeral3} presents the improved beam search decoding algorithm.
Section \uppercase\expandafter{\romannumeral4}
shows the compression of a dictionary used in the beam search decoding.
In Section \uppercase\expandafter{\romannumeral5},
we implement the fixed-point CTC-decoder.
Section \uppercase\expandafter{\romannumeral6} concludes this paper.

\section{Background}
\subsection{Review of CTC}
Assume that the output sequence and the target sequence of the system shown in Fig. \ref{fig:system framework} have $K$ labels, and another blank label \o~is covered in the intermediate calculations. The \o~means a null emission. Define $X=({X_1},...,{X_T})$ as the input sequence of the network.
Define $Y=({Y_1},...,{Y_T})$ as the output sequence of the network. At time $t$, we have ${Y_t}=({Y_t^1},...,{Y_t^{K+1}})$. So each of the outputs of the softmax layer represents the probability of each label:

\begin{equation}\label{eq:pr_kt_x}
Pr(k,t|X)=\frac{exp({Y_t^k})}{ \sum_{i=1}^{K+1}exp({Y_t^i})}.
\end{equation}

A CTC path $\pi$ which is introduced in \cite{graves2006connectionist} as a sequence of labels (including \o), can be expressed as $\pi=({\pi_1},...,{\pi_T})$.
Assuming that the probabilities of emitting a label at different times are conditionally independent, the probability of a CTC path $\pi$ can be calculated as follows:
\begin{equation}\label{eq:pr_pi_x}
Pr(\pi|X)= \prod_{t=1}^{T}Pr({\pi_t},t|X).
\end{equation}

The target sequence L is corresponding to a set of CTC paths, and the mapping function $\beta$ is described in \cite{graves2006connectionist}.
The function $\beta$ removes all repeated labels and blanks from the path (e.g.~$\beta(c~\phi~\phi~a~\phi~ t)=\beta(c~c~\phi~a~a~a~\phi~\phi~t~ t)=cat$).
We can evaluate the probability of the target sentence as the sum of the probabilities of all the CTC paths in the set:
\begin{equation}\label{eq:pr_L_x}
Pr(L|X)= \sum_{\pi\in\beta^{-1}(L)}Pr(\pi|X).
\end{equation}

However, it is virtually impossible to sum the probabilities of all the paths in $\beta^{-1}(L)$.
To calculate $Pr(L|X)$, the CTC Forward-Backward Alogrithm was invented in \cite{graves2006connectionist}.
Afterwards, the network can be trained with the CTC objective function:
\begin{equation}\label{eq:CTC_x}
CTC(X)=-log Pr(L|X).
\end{equation}

\subsection{CTC Beam Search Decoding}
Decoding a CTC network means finding the most probable output sequence for a given input.
The simplest way to decode it is the best path decoding introduced in \cite{graves2006connectionist}:
by picking the single most probable label at every time step, the most probable sequence will correspond to the most probable labelling.
Some works use this decoding method to build the CTC-layers in their hardware architectures of RNNs \cite{rybalkin2017hardware}.
Although this way can already provide useful transcriptions, its limited accuracy is not sufficient to meet the demands of many sequence tasks\cite{Zenkel2017Comparison}.

The CTC beam search decoding searches for the most probable sequence in all the sequences ($length \leq T$)
combined with $K$ labels (\o~will not appear in output sequence).
The number of all the sequences is growing exponentially with the increase of T,
but the number of the sequences searched with the CTC beam search decoding is no larger than $K\cdot W \cdot T$.
The beam width $W$ determines the complexity and accuracy of the algorthm.
If $W$ is big enough, the probability will be one so that the beam search is equal to the breadth first search (BFS).
However, the algorithm will be too complex.
But if $W$ is too small, the probability of using beam search to find the correct answer will be too small.
So there is a trade off between the size of $W$ and the accuracy.

The probability of output sequence $\boldsymbol{y}$ (including \o) at time $t$ is defined as $Pr(\boldsymbol{y},t$).
All the paths in $\beta^{-1}(\boldsymbol{y})$ can be classified into two sets, $\xi_1(\boldsymbol{y})$ and $\xi_2(\boldsymbol{y})$.
The last label of any path in $\xi_1(\boldsymbol{y})$ must be \o, while the last label of any path in $\xi_2(\boldsymbol{y})$ can be any label except \o.
Defining the sum of the probabilities of the paths in $\xi_1(\boldsymbol{y})$ and $\xi_2(\boldsymbol{y})$ as $Pr^{-}(\boldsymbol{y},t)$ and $Pr^{+}(\boldsymbol{y},t)$,
respectively, we have $Pr(\boldsymbol{y},t)=Pr^{-}(\boldsymbol{y},t)+Pr^{+}(\boldsymbol{y},t).$ Define $\theta$ as the empty sequence  ($Pr^+(\theta,t)=0$), $\boldsymbol{\hat{y}}$ as the prefix of $\boldsymbol{y}$ with its last label removed, and $\boldsymbol{y^e}$ as the last label of $\boldsymbol{y}$.
The CTC beam search decoding is described in Algorithm \ref{alg:CTC Beam Search Decoding}.

\begin{algorithm}
  \caption{CTC Beam Search Decoding}
  \label{alg:CTC Beam Search Decoding}
  \begin{algorithmic}[1]
  \STATE $t=0$
  \STATE $B \gets \{  {\theta}  \}$, $Pr^{-}(\theta,t) \gets 1$
  \WHILE {$t<T$}
  \STATE $\hat{B} \gets$ the $W$ most probable sequences in B
  \STATE $B \gets \{ \}$
  \FOR {$\boldsymbol{y} \in \hat{B}$}
  \STATE $Pr^{-}(\boldsymbol{y},t) \gets Pr(\boldsymbol{y},t-1)Pr(\phi,t|X)$
  \IF {$y \neq \theta$}
  \STATE $Pr^{+}(\boldsymbol{y},t) \gets Pr^{+}(\boldsymbol{y},t-1)Pr(\boldsymbol{y^e},t|X)$
  \IF {$\boldsymbol{\hat{y}} \in \hat{B}$}
  \STATE $Pr^{+}(\boldsymbol{y},t) \gets Pr^{+}(\boldsymbol{y},t) + Pr(\boldsymbol{y^e},
        \boldsymbol{\hat{y}},t)$
  \ENDIF
  \ENDIF
  \STATE $Pr(\boldsymbol{y},t)=Pr^{+}(\boldsymbol{y},t)+Pr^{-}(\boldsymbol{y},t)$,
      Add $\boldsymbol{y}$ to $B$
  \FOR {$k=1...K$}
  \STATE $Pr^{-}(\boldsymbol{y}+k,t) \gets 0$
  \STATE $Pr^{+}(\boldsymbol{y}+k,t) \gets Pr(k,\boldsymbol{y},t)$
  \STATE $Pr(\boldsymbol{y}+k,t)=Pr^{-}(\boldsymbol{y}+k,t)+Pr^{+}(\boldsymbol{y}+k,t)$
  \STATE Add $\boldsymbol{y}+k$ to $B$
  \ENDFOR
  \ENDFOR
  \STATE $t \gets t+1$
  \ENDWHILE
  \STATE output the most probable sequence in $\hat{B}$
  \end{algorithmic}
  \end{algorithm}

$Pr(k,t|X)$ is defined in Equation (\ref{eq:pr_kt_x}). The extension probability $Pr(k,\boldsymbol{y},t)$ is defined in Equation (\ref{eq:pr_kyt}):

\begin{equation}\label{eq:pr_kyt}
Pr(k,\boldsymbol{y},t)=
\begin{cases}
Pr(k|\boldsymbol{y})Pr(k,t|X)Pr^{-}(\boldsymbol{y},t-1)& \text{${y^e}=k,$}\\
Pr(k|\boldsymbol{y})Pr(k,t|X)Pr(\boldsymbol{y},t-1)& \text{${y^e}\neq k.$}
\end{cases}
\end{equation}

The transition probability from $\boldsymbol{y}$ to $\boldsymbol{y}+k$ is $Pr(k|\boldsymbol{y})$,
allowing prior linguistic information to be integrated.
All $Pr(k|\boldsymbol{y})$ are set by the LM. If no LM is used, all $Pr(k|\boldsymbol{y})$ are set to 1.
If the LM is just a dictionary,
$Pr(k|\boldsymbol{y})$ will be set in accordance with Equation (\ref{eq:pr_k_y}).
\begin{equation}\label{eq:pr_k_y}
    Pr(k|\boldsymbol{y})=
    \begin{cases}
    1& \text{$(\boldsymbol{y}+k)$ is in the dictionary,}\\
    0& \text{$(\boldsymbol{y}+k)$ is not in the dictionary.}
    \end{cases}
\end{equation}

If a more complicated LM is used, $Pr(k|\boldsymbol{y})$ will be set differently,
which has been discussed in \cite{Graves2014Towards}.
In this work we just focus on the dictionary LM.

\subsection{Low-Complexity Softmax Function}
The softmax function is described in Equation (\ref{eq:pr_kt_x}),
which is widely used in various neural network systems.
Our previous work \cite{wang2018high} proposed a high-speed and low-complexity
architecture for softmax function.
For computational characteristics of CTC-decoder, a variant model for softmax is used
in this work.

\subsubsection{Log-Sum-Exp Trick}
The log-sum-exp trick is adopted as Equation (\ref{eq:log-sum-exp})\cite{Yuan2017Efficient}.
After this mathematical transformation, we not only replace the division operation by a
subtraction operation but also avoid numerical underflow.

\begin{equation}\label{eq:log-sum-exp}
	\begin{split}
		\begin{aligned}
			p_k &= \frac{y_k-y_{max}}{\sum_{i=1}^{K+1}exp(y_i-y_{max})} \\
				&= exp(y_k-y_{max} - ln(\sum_{i=1}^{K+1}exp(y_i-y_{max}))) \\
				&(\forall k \in {1,2,...,K+1}, y_{max}\geq y_k).
		\end{aligned}
	\end{split}
\end{equation}

\subsubsection{The Transformation of Exponential Function}
The exponential function is not so easy to calculate, but if we limit its inputs within a specific range,
the calculation will be much simplified.

Transform $e^{y_i}$ with the following expression:
\begin{equation}\label{eq:exp}
	\begin{split}
		\begin{aligned}
			e^{y_i} &=2^{y_i\cdot\log_{2}e} = 2^{u_i+v_i}=(2^{u_i})\cdot(2^{v_i}).  \\
				u_i &= \left \lfloor y_i\cdot \log_{2}e \right \rfloor ,
				~~~ v_i = y_i - u_i .
		\end{aligned}
	\end{split}	
\end{equation}
Since $u_i$ is an integer, and $v_i$ is limitd in $(0, 1]$, we can replace the original exponential unit with the operation $f(v_i)=2^{v_i}$ and a simple shift operation.
The operation $f(v_i)=2^{v_i}$ can be approximated as functions $f(x)=x+d_1$
or $f(x)=x+d_2$, where two bias values $d_1$ and $d_2$ correspond to the first and Second
exponential operations, respectively.

\subsubsection{The Transforamtion of Logarithmic Function}
Similarly, we can simplify the calculation of logarithmic function by limiting the range of its input.

Transform $lnF$ with the following expression:
\begin{equation}\label{eq:log}
	\begin{split}
		\begin{aligned}
			lnF &= ln2\cdot log_{2}F = ln2\cdot (\omega + log_{2}\kappa). \\
			\omega &=  \left \lfloor log_{2}F \right \rfloor,
			~~ \kappa = F \div 2^{\omega} .
		\end{aligned}
	\end{split}	
\end{equation}
As a result, $\kappa$ is limited in $[1,2)$, so the approximation $log_{2}\kappa \approx k-1$ can be used. Finally, the logarithmic function can be simplified as $lnF=ln2\cdot (\kappa - 1 + \omega)$.

\section{CTC Beam sarch decoding improvements}
This section improves the CTC beam search decoding (Algorithm \ref{alg:CTC Beam Search Decoding}) to save memory space.
Additionally, the time complexity of the improved algorithm (Algorithm \ref{alg:all improvement}) is the same as that of Algorithm \ref{alg:CTC Beam Search Decoding}, which is $O(T\cdot W\cdot K)$. As mentioned before, the improved serial algorithm is suitable for both software and hardware decoding.

\subsection{Memory Space Required by Original CTC Beam Search Decoding Algorithm}
\begin{figure}[ht]
\centering
\includegraphics[scale=0.8]{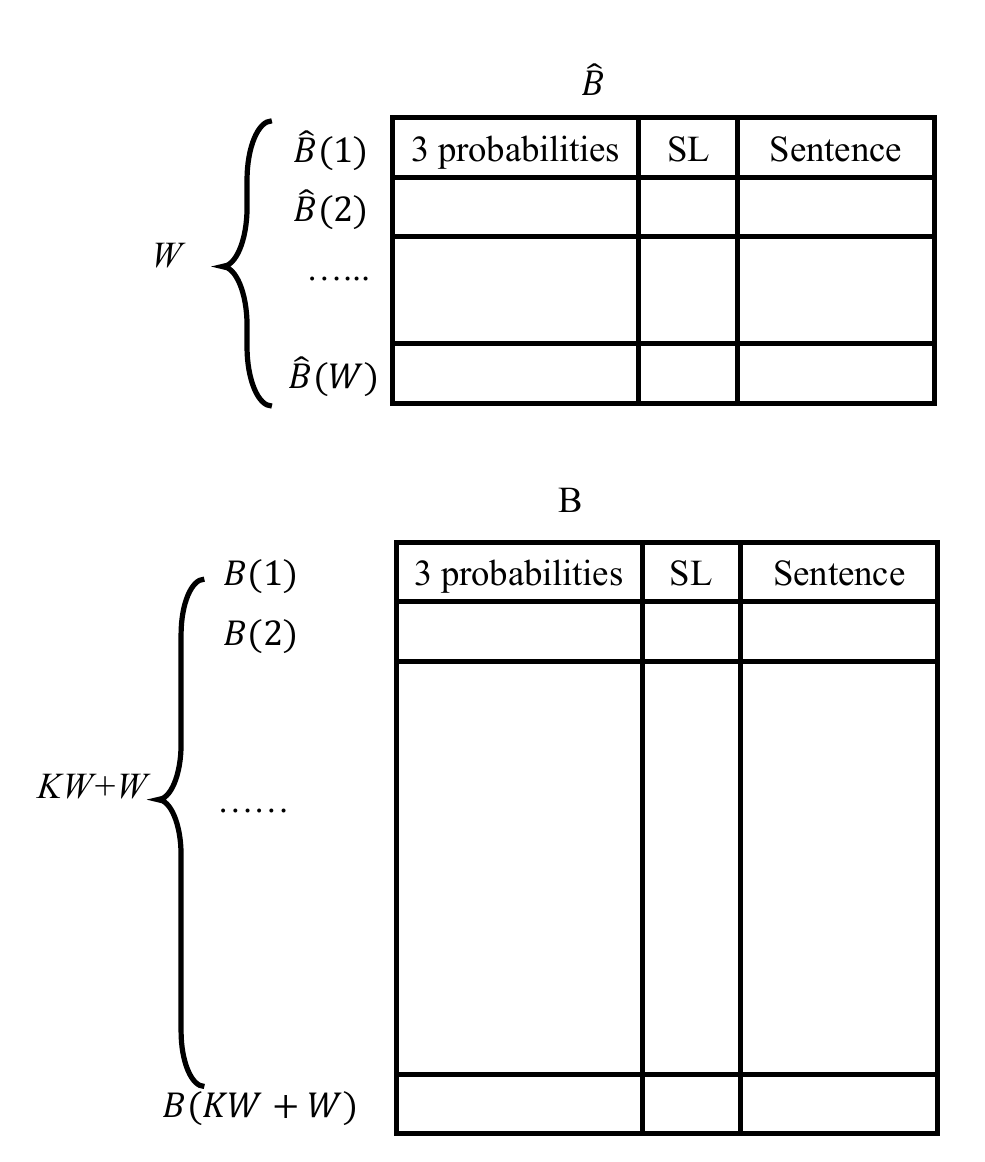}
\caption{The storage structure of the Algorithm \ref{alg:CTC Beam Search Decoding}. The width of each probability is determined by the experiment. SL is used to store some necessary information related to LM. Its width is decided by the LM. Each Sentence consists of $T$ labels, requiring $T\cdot \lceil \log_2K \rceil$ bits.}
\label{fig:origin storage}
\end{figure}

The storage structure of the original algorithm (Algorithm \ref{alg:CTC Beam Search Decoding}) is described in Fig. \ref{fig:origin storage}.
There are $(K+2)W$ label sequences. $W$ sequences are in $\hat{B}$, and $(K+1)W$ sequences are in $B$.
$\hat{B}(i)$ or $B(i)$ represents each label sequence in $\hat{B}$ or $B$.

For convenience of discussion, we use $\boldsymbol{y}$ to denote a $\hat{B}(i)$ or a $B(i)$.
To store each $\boldsymbol{y}$, required information includes the three probabilities ($Pr^-(\boldsymbol{y},t),Pr^+(\boldsymbol{y},t),Pr(\boldsymbol{y},t)$), the SL (used to store some necessary information related to LM) and the Sentence. The Sentence is used to store every label of $\boldsymbol{y}$ in chronological order. Considering the worst situation, each Sentence consists of $T$ labels.

\subsection{First Improvement: Decrease the Number of Sequences in $\boldsymbol{B}$}
Most of the sequences stored in $B$ are useless. In fact, only $W$ of the sequences in $B$ can be reserved in each iteration. In this subsection, the number of sequences in $B$ is reduced to $W$ sequences.

We define the minimum of $Pr(B(i),t)$ as $min(Pr)$.
The solution of working out $min(Pr)$ can be a sorting block on hardware platform,
or using a min-heap\cite{Cormen2001Introduction}. The min-heap is a binary tree,
and each node represents each  $Pr(B(i),t)$. The value of the root node is $min(Pr)$,
and the value of each node other than the root node is not less than its parent node.
When a new sequence is evaluated, its probability will be compared with $min(Pr)$. Only if the probability is larger than $min(Pr)$, the new sequence can take place of the sequence whose probability is $min(Pr)$.

Nevertheless, simply reducing the size of $B$ and giving the $min(Pr)$ could not give the right answer as Algorithm \ref{alg:CTC Beam Search Decoding} gives. Pay attention to the line 11 of Algorithm \ref{alg:CTC Beam Search Decoding}, where a special situation is taken into consideration. For example, assume $\boldsymbol{\hat{y}}=(a~b~c)$ and $\boldsymbol{y}=(a~ b~c~d)$. Calculating $Pr^+(\boldsymbol{y},t)$ requires $Pr(d,\boldsymbol{\hat{y}},t)$, which has probably been discarded if it is smaller than $min(Pr)$. Therefore,  all values of $Pr(\boldsymbol{y}+k,t)$ must be calculated before the values of $Pr(\boldsymbol{y},t)$ in the improved algorithm, and then special probabilities like $Pr(d,\boldsymbol{\hat{y}},t)$ can be reserved. We use three arrays, $B1$, $B2$ and $B3$, to save these probabilities. The detailed descriptions of them are listed in TABLE \ref{table:var in alg2}

\newcommand{\tabincell}[2]{\begin{tabular}{@{}#1@{}}#2\end{tabular}}
\begin{table}
  \normalsize
  \centering
  \caption{}
  \label{table:var in alg2}
  \begin{tabular}{|l|l|}
  \hline
  Variable&  Description  \\
  \hline
  $B1(i) $& the index of prefix of $\hat{B}(i).Sentence$ \\
          & if the prefix exists in $\hat{B}$\\

  $B2(i)$ &  the last label $k$ of $\hat{B}(i).Sentence$  \\

  $B3(i)$ &  $Pr(B2(i), \hat{B}(B1(i)), t )$ \\
  \hline
  \end{tabular}
\end{table}

The modified algorithm is Algorithm \ref{alg:First improvement}. $B$ can be set as a min-heap, and finding $min(Pr)$ will be very easy (the position of it will be fixed in $B$). But the heap needs to be adjusted when new elements come in. Another choice is to figure out the position of $min(Pr)$ in real time, which will be easy to implement on hardware platform by using a sorting block.

\begin{algorithm}
\caption{CTC Beam Search Decoding with First Improvement}
\label{alg:First improvement}
\begin{algorithmic}[1]
\STATE $t\leftarrow0$
\STATE
$\hat{B}(1).Sentence\leftarrow\theta$,$Pr^-(\hat{B}(1))\leftarrow1$
\WHILE {$t<T$}
\FOR {$(\hat{B}(i),\hat{B}(j))\in\hat{B},(i\neq j)$}
\IF {$\hat{B}(i).Sentence=\hat{B}(j).Sentence+k$}
\STATE $B1(i)=j,B2(i)=k$
\ENDIF
\ENDFOR
\FOR {$\hat{B}(i)~in~\hat{B}$}
\FOR {$k=1...K$}
\STATE $Temp\leftarrow Pr(k,\hat{B}(i),t)$
\STATE $T_S\leftarrow information~received~from~LM$
\IF {$(\hat{B}(i)=B1(j))\bigwedge(k=B2(j))$}
\STATE $B3(j)\leftarrow Temp$
\ENDIF
\STATE $find~B(mi)~as~min(Pr)~:~\forall j~\neq~ mi$,     $Pr(B(mi),t)\leq Pr(B(j),t)$
\IF {$Temp > min(Pr)$}
\STATE $Pr(B(mi),t)\leftarrow Temp$
\STATE $B(mi).SL\leftarrow T_S$
\STATE $Pr^+(B(mi),t)\leftarrow Temp$
\STATE $Pr^-(B(mi),t)\leftarrow 0$
\STATE $B(mi).Sentence\leftarrow \hat{B}(i).Sentence+k$
\STATE $if~B~is~a~$min-heap$,~adjust~it$
\ENDIF
\ENDFOR
\ENDFOR
\FOR {$\hat{B}(i)~in~\hat{B}$}
\STATE $Temp^-\leftarrow Pr(\hat{B}(i),t-1)\cdot Pr(\phi,t|X)$
\STATE $Temp^+\leftarrow Pr^+(\hat{B}(i),t-1)\cdot Pr({\hat{B}(i)}^e,t)+B3(i)$
\STATE $Temp\leftarrow Temp^- + Temp^+$
\IF {$\hat{B}(i).Sentence=B(j).Sentence $}
\STATE $(Pr^-(B(j),t),Pr^+(B(j),t),Pr(B(j),t))$ $\leftarrow(Temp^-,Temp^+,Temp)$
\STATE
$B(j).SL\leftarrow\hat{B}(i).SL$
\STATE $if~B~is~a~$min-heap$,~adjust~it$
\ELSE
\STATE $find~B(mi)~as~min(Pr)$ (same as line 16)
\IF {$Temp > min(Pr)$}
\STATE $Pr(B(mi),t)\leftarrow Temp$
\STATE $B(mi).SL\leftarrow\hat{B}(i).SL$
\STATE $Pr^+(B(mi),t)\leftarrow Temp^+$
\STATE $Pr^-(B(mi),t)\leftarrow Temp^-$
\STATE $B(mi).Sentence\leftarrow \hat{B}(i).Sentence$
\STATE $if~B~is~a~$min-heap$,~adjust~it$
\ENDIF
\ENDIF
\ENDFOR
\STATE $\hat{B}\leftarrow B$
\STATE $t \gets t+1$
\ENDWHILE
\STATE output the most probable sequence in $\hat{B}$
\end{algorithmic}
\end{algorithm}

\begin{figure}[ht]

  \centering
  \includegraphics[scale=0.85]{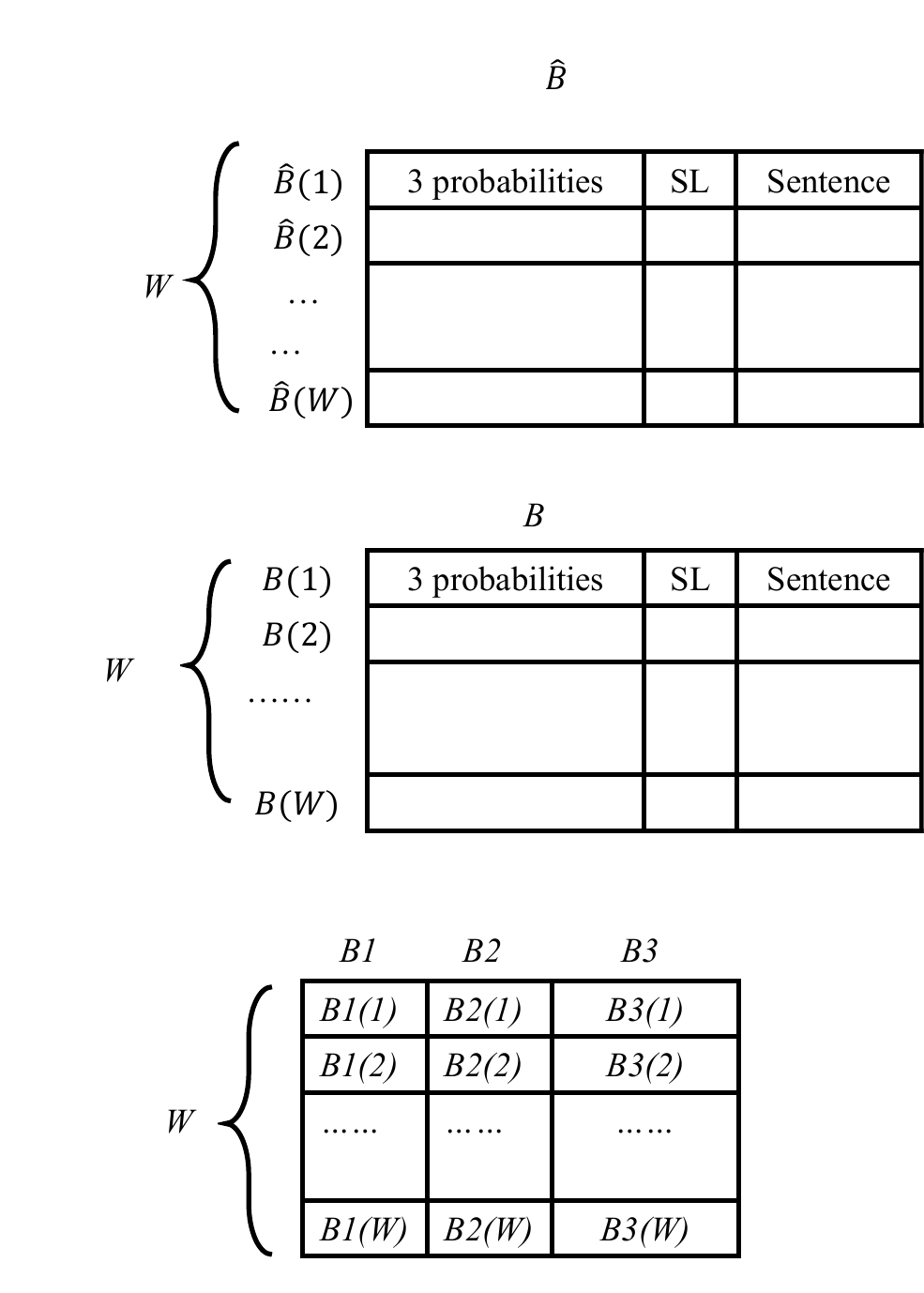}
  \caption{The storage structure of the Algorithm \ref{alg:First improvement}. The width of B1 is $\lceil \log_2W \rceil$. The width of B2 is $\lceil \log_2K \rceil$. The width of B3 is the same as the width of each probability in $B$. }
  \label{fig:first storage}
  \end{figure}

Algorithm \ref{alg:First improvement} obviously outperforms Algorithm \ref{alg:CTC Beam Search Decoding}.
Firstly, Algorithm \ref{alg:First improvement} solves the problem of finding the $W$ most probable sequences in $B$,
which is mentioned in the line 4 of Algorithm \ref{alg:CTC Beam Search Decoding}. Secondly, the memory space used in Algorithm \ref{alg:First improvement} is obviously smaller than Algorithm \ref{alg:CTC Beam Search Decoding}.
The storage structure of Algorithm \ref{alg:First improvement} is shown in Fig. \ref{fig:first storage}.
The memory space required by $B1,B2$ and $B3$ is much smaller than $B$, and now the size of $B$ is the same as $\hat{B}$. In most cases, the reduction of B will compress the required memory space to nearly $\frac{3}{K+2}$ of the original size. $K$ is probably much larger than 10, so the compression ratio will be much larger than 5. Thirdly, the number of assignments of $B$ is significantly reduced.
Compared with Algorithm \ref{alg:CTC Beam Search Decoding}, Algorithm \ref{alg:First improvement} takes a few more steps to fill $B1$ and $B2$, which is a perfectly acceptable tradeoff.

\subsection{Second Improvement: Remove All the Sentences in $B$ }
Although Algorithm \ref{alg:First improvement} has saved most of the required memory space, there is still redundant storage. In this subsection, we remove all the Sentences of $B$ to further improve the beam search algorithm and get a higher compression ratio.

In Fig. \ref{fig:first storage}, $\hat{B}.Sentence$ and $B.Sentence$ take up a lot of space. Each $B(i)$ or $\hat{B}(i)$ contains only three probabilities, but each $B(i).Sentence$ or $\hat{B}(i).Sentence$ has hundreds of labels (in most cases, $T$ is much larger than 100). The width of each probability saved in $B$ or $\hat{B}$ is probably no bigger than 64 bits. The width of each label is $\lceil \log_2K \rceil$. If $K$ is bigger than 10, the space used by $B.Sentence$ will almost be twice the size of the space used by all the probabilities in $B$.

The only function of $B.Sentence$ is to iterate and update $\hat{B}.Sentence$, as shown in the line 47 of Algorithm \ref{alg:First improvement}. However, $\hat{B}.Sentence$ can be iterated and updated without $B.Sentence$. The prefix of $B(i).Sentence$ with its last label removed or the $B(i).Sentence$ itself can certainly be found in $\hat{B}.Sentence$, by mapping sequences in $B$ into sequences in $\hat{B}$. Define this mapping as $\rho:B\rightarrow\hat{B}$. $\rho$ is a general mapping, which means sequences in $\hat{B}$ may have no preimage or more than one preimages. Based on $\rho$, Algorithm \ref{alg:Update} is introduced to replace line 47 in Algorithm \ref{alg:First improvement}.
New arrays $A1$, $A2$, $d$ and $c$ are defined in TABLE \ref{table:var in alg3}.
The size of A1 is the same as B1, and the size of A2 is as big as B2. Boolean arrays $d$ and $c$ only consume $2W$ bits. Note that an LOD can be reused on hardware platform for the calculation in the line 22 of Algorithm \ref{alg:Update}.

\begin{table}
  \normalsize
  \centering
  \caption{}
  \label{table:var in alg3}
  \begin{tabular}{|l|l|}
  \hline
  Variable&  Description  \\
  \hline
  $A1(i) $& the index of the prefix of $B(i).Sentence$  \\
          &  or $B(i).Sentence$ itself in $\hat{B}$    \\

  $A2(i)$ & the last label $k$ of $B(i).Sentence$  or $zero$ \\

  $d(i)$  & whether the information in $\hat{B}(i)$  has been \\
          & updated by $B$\\
  $c(i)$  & whether $B(i)$ has replaced the information \\
          & in $\hat{B}$  \\
  \hline
  \end{tabular}
\end{table}

\begin{algorithm}

\caption{Update $\hat{B}$ without B.Sentence}
\label{alg:Update}
\begin{algorithmic}[1]
\FOR {$i=1...W$}
\STATE $d(i)\leftarrow false,c(i)\leftarrow false$
\STATE $A1(i)\leftarrow \rho(B(i))$
\IF {when $B(i)$ was added in $B$, the Sentence was enlarged with $k$}
\STATE $A2(i)\leftarrow k $
\ELSE
\STATE $A2(i)\leftarrow 0 $
\ENDIF
\ENDFOR
\FOR {$i=1...W$}
\IF {($d(A1(i))=false$)}
\STATE $\hat{B}(A1(i)).probability$\&$SL \leftarrow B(i)$
\IF {$A2(i)>0$}
\STATE $\hat{B}(A1(i)).Sentence=\hat{B}(A1(i)).Sentence+A2(i)$
\ENDIF
\STATE $d(a(i))=true$
\STATE $c(i)=true$
\ENDIF
\ENDFOR
\FOR {$i=1...W$}
\IF {$c(i)=false$}
\STATE $j\leftarrow the~leading~0~in~d$
\STATE $\hat{B}(j).probability$\&$SL \leftarrow B(i)$
\STATE $\hat{B}(j).Sentence\leftarrow \hat{B}(i).Sentence$
\IF {$A2(i)>0$}
\STATE $\hat{B}(i).Sentence=\hat{B}(i).Sentence+A2(i)$
\ENDIF
\ENDIF
\ENDFOR
\end{algorithmic}
\end{algorithm}

The key problem solved by Algorithm \ref{alg:Update} can be outlined as follows:

\subsubsection{The Problem}Define $C_p$ as a combination of $W$ numbers in $\{1,2,...,W\}$ (not ordered). $C_p$ is saved in array $\bar{L}$ (ordered). A $W$-length array is defined as $L$, $L=(C^1,C^2,...,C^W)$. Define $Comb(L)$ as a combination of all the superscripts of $C$ in $L$. The purpose is to transform $L$ so that $Comb(L)$ is equal to $C_p$, using only one operation: copying its own element to cover another element of it. Apart from $L$, there is no other place to store any $C^i$. Meanwhile, try to make the number of the copies as few as possible.

\subsubsection{An Example}Shown in TABLE \ref{table2}.

\begin{table}[H]
\normalsize
  \centering
  \caption{}
  \label{table2}
\begin{tabular}{cc}
\hline
  Name   &     Value\\
\hline
W& 8\\
$L$ ($in~beginning$)& $(C^1,C^2,C^3,C^4,C^5,C^6,C^7,C^8)$\\
$Comb(L)$     & (1, 2, 3, 4, 5, 6, 7, 8)           \\
$C_p$         & (4, 6, 8, 6, 3, 3, 7, 1)        \\
\hline
\end{tabular}
\end{table}

\subsubsection{The Solution}
In Algorithm \ref{alg:Update}, the first loop is the initialization, and the main procedure is comprised of the rest two loops.
In the first loop of the main procedure, the $C^j$ in correct place is fixed. In this example,
we have $d=(true,false,true,true,false,true,true,true)$ and $c=(true,true,true,false,true,false,true,true)$ after the first loop of the main procedure.
After the main procedure, $L$ is transformed to what we want:
$( C^1,C^6,C^3,C^4,C^3,C^6,C^7,C^8)$.

\begin{figure}[ht]

\centering
\includegraphics[scale=0.85]{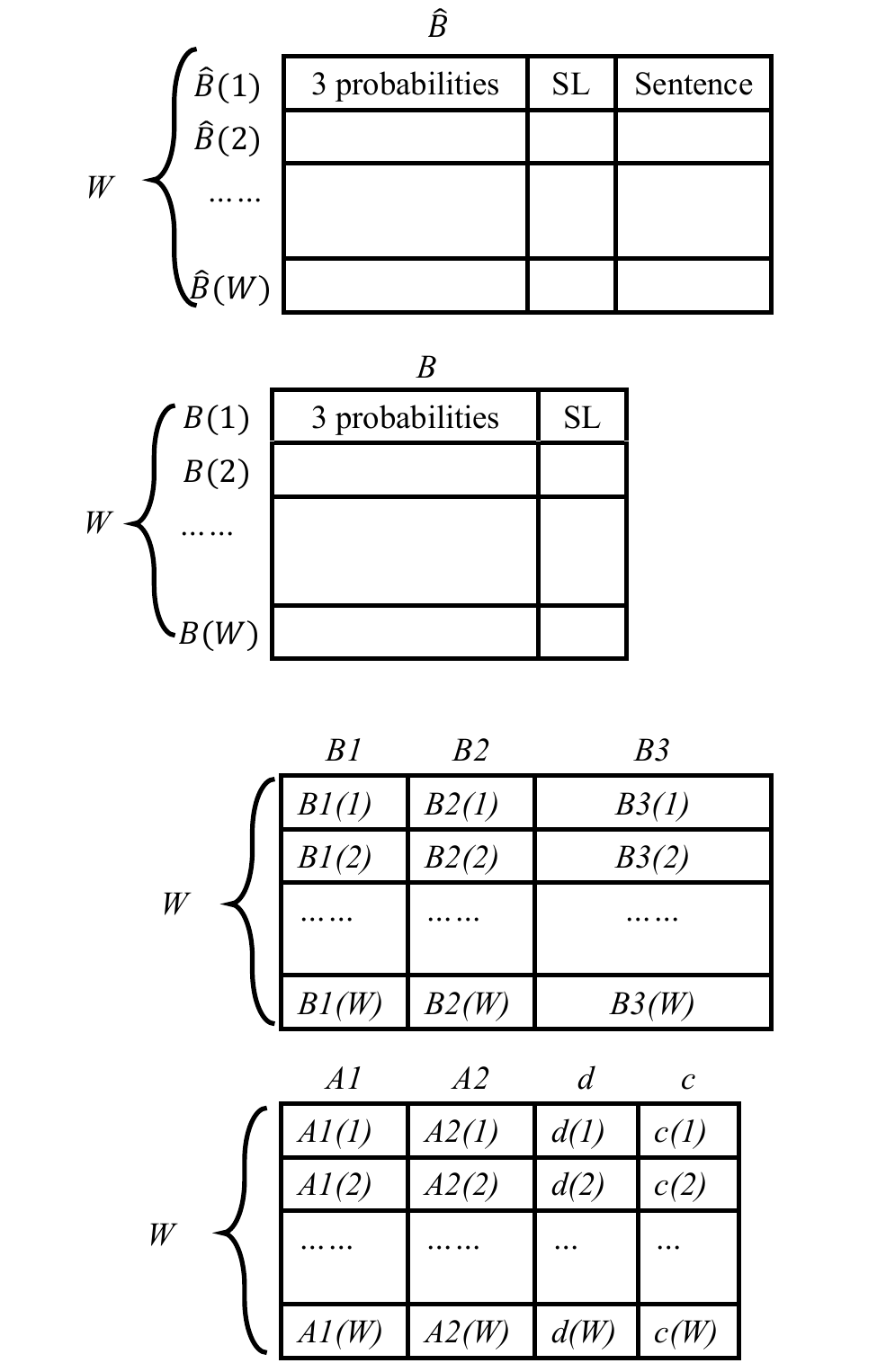}
\caption{The storage structure of the Algorithm \ref{alg:all improvement}. The width of A1 is the same as B1. The width of A2 is the same as B2. The widths of $c$ and $d$ are both 1 bit. }
\label{fig:final storage}
\end{figure}

On one hand, Algorithm \ref{alg:Update} keeps the number of the assignments of $\hat{B}.Sentence$ as few as possible. On the other hand, Algorithm \ref{alg:Update} removes all the Sentences of $B$, but it needs more space for $A1,A2,d$ and $c$. However, the size of them is far smaller than $B.Sentence$, which means Algorithm \ref{alg:Update} further compresses the memory space used by the beam search decoding. The remaining memory space after these first two improvements can be seen in Fig. \ref{fig:final storage}.

\subsection{Third Improvement: Prevent Probabilities from Being Too Small}
The first two improvements have already made the CTC beam search decoding highly memory-efficient, but we find all the probabilities in $\hat{B}$ tend to become smaller in decoding. Because the output of softmax is smaller than 1, all the probabilities will converge to 0. To tackle this problem, an adjustment of these probabilities is added after the update of $\hat{B}$. Here we give a conclusion which is proved in Appendix A: At each time after the update of $\hat{B}$, if all the probabilities of $\hat{B}$ increase (or decrease) by the same times, the output of the whole system will not change.

According to this conclusion, a lower limit (named as $P_l$) is set for the maximum of $Pr(\hat{B}(i),t)$ (named as $max(Pr)$). After the update of $\hat{B}$, $max(Pr)$ is compared with $P_l$. If $max(Pr)$ is less than $P_l$, it will be enlarged to ensure that it is no smaller than $P_l$. The last step is to increase all the rest probabilities (including $Pr^-$,$Pr^+$ and $Pr$) by the same scale. To make the algorithm easier to be implemented in hardware we use Equation (\ref{eq:pl}) to determine $P_l$.

\begin{equation}\label{eq:pl}
\frac{1}{4W}<P_l\leq\frac{1}{2W}, P_l=2^n, n\leq-1 \bigwedge n \in Z.
\end{equation}

As a fix-pointed binary number, only one bit of $P_l$ is set to 1. The position of this 1 is called as $index(P_l)$. The calculation steps of this adjustment are shown in Algorithm \ref{alg:Adjust}. This algorithm also guarantees that $\sum_{i=1}^{W}Pr(\hat{B}(i),t)<1$.

\begin{algorithm}
\caption{Adjust Probabilities}
\label{alg:Adjust}
\begin{algorithmic}[1]
\STATE find $\hat{B}(mi)$ as $max(Pr)$ :$\forall j~\neq~ mi, Pr(\hat{B}(mi),t)\geq Pr(\hat{B}(j),t)$
\STATE $j\leftarrow$ position of the leading 1 in $max(Pr)$
\IF {$j<index(P_l)$ ,($max(Pr)<P_l$)}
\STATE $i\leftarrow index(P_l)-j$
\ENDIF
\FOR {all probabilities in $\hat{B}$}
\STATE probability=probability$<<i$
\ENDFOR
\end{algorithmic}
\end{algorithm}

Again, the LOD can be reused for the step in the line 2. The sorting block for finding the maximum can also be reused in the line 49 of Algorithm \ref{alg:all improvement}. As a result, this algorithm consumes few resources on hardware platform, but solves the problem of probabilities in $\hat{B}$ being too smaller.

\begin{algorithm}
\caption{CTC Beam Search Decoding with All Improvements}
\label{alg:all improvement}
\begin{algorithmic}[1]
\STATE $t\leftarrow0$
\STATE
$\hat{B}(1).sentence\leftarrow\theta$,$Pr^-(\hat{B}(1))\leftarrow1$
\WHILE {$t<T$}
\FOR {$(\hat{B}(i),\hat{B}(j))\in\hat{B},(i\neq j)$}
\IF {$\hat{B}(i).sentence=\hat{B}(j).sentence+k$}
\STATE $B1(i)=j,B2(i)=k$
\ENDIF
\ENDFOR
\FOR {$\hat{B}(i)~in~\hat{B}$}
\FOR {$k=1...K$}
\STATE $Temp\leftarrow Pr(k,\hat{B}(i),t)$
\STATE $T_S\leftarrow information~received~from~LM$
\IF {$(\hat{B}(i)=B1(j))AND(k=B2(j))$}
\STATE $B3(j)\leftarrow Temp$
\ENDIF
\STATE $find~B(mi)~as~min(Pr)~:~\forall j~\neq~ mi$,     $Pr(B(mi),t)\leq Pr(B(j),t)$
\IF {$Temp > min(Pr)$}
\STATE $Pr(B(mi),t)\leftarrow Temp$
\STATE $B(mi).SL\leftarrow T_S$
\STATE $Pr^+(B(mi),t)\leftarrow Temp$
\STATE $Pr^-(B(mi),t)\leftarrow 0$
\STATE $A1(mi)\leftarrow i$
\STATE $A2(mi)\leftarrow k$
\STATE $if~B~is~a~$min-heap$,~adjust~it$
\ENDIF
\ENDFOR
\ENDFOR
\FOR {$\hat{B}(i)~in~\hat{B}$}
\STATE $Temp^-\leftarrow Pr(\hat{B}(i),t-1)\cdot Pr(\phi,t|X)$
\STATE $Temp^+\leftarrow Pr^+(\hat{B}(i),t-1)\cdot Pr({\hat{B}(i)}^e,t)+B3(i)$
\STATE $Temp\leftarrow Temp^- + Temp^+$
\IF {$B1(i)=A1(j)\bigwedge B2(i)=A2(j)$}
\STATE $(Pr^-(B(j),t),Pr^+(B(j),t),Pr(B(j),t))$  $\leftarrow(Temp^-,Temp^+,Temp)$
\STATE $B(j).SL\leftarrow\hat{B}(i).SL$
\STATE $if~B~is~a~$min-heap$,~adjust~it$
\ELSE
\STATE $find~B(mi)~as~min(Pr)$ (same as line 16)
\IF {$Temp > min(Pr)$}
\STATE $Pr(B(mi),t)\leftarrow Temp$
\STATE $B(mi).SL\leftarrow\hat{B}(i).SL$
\STATE $Pr^+(B(mi),t)\leftarrow Temp^+$
\STATE $Pr^-(B(mi),t)\leftarrow Temp^-$
\STATE $A1(mi)\leftarrow i$
\STATE $A2(mi)\leftarrow k$
\STATE $if~B~is~a~$min-heap$,~adjust~it$
\ENDIF
\ENDIF
\ENDFOR
\STATE Update $\hat{B}$ without $B.sentence$(Algorithm \ref{alg:Update})
\STATE Adjust Probabilities (Algorithm \ref{alg:Adjust})
\STATE $t \gets t+1$
\ENDWHILE
\STATE output the most probable sequence in $\hat{B}$
\end{algorithmic}
\end{algorithm}

\section{Compressed Dictionary}
This section talks about the LM visitor module and the LM stored in memory.
An LM is integrated to improve the precision of decoding by adjusting the value of $Pr(k|y)$ in Equation (\ref{eq:pr_kyt}). In Algorithm \ref{alg:all improvement}, the calculation of $Pr(k|y)$ is only required in the line 11, where $y=\hat{B}(i)$. The dictionary is the simplest LM, including a specific number of words. In this section, an English dictionary (191,735 words, from the vocabulary of OpenSLR) is used as an example to demonstrate the effect of the compression. Subsection A talks about the basic data structure (DS) of the dictionary. In Subsection B and C, strategies of the compression are explained. In Subsection D, an algorithm is presented to apply the compressed dictionary to decoding.

\subsection{Basic Data Structure: Trie-tree}
\begin{figure}[ht]
  \centering
  \includegraphics[scale=0.70]{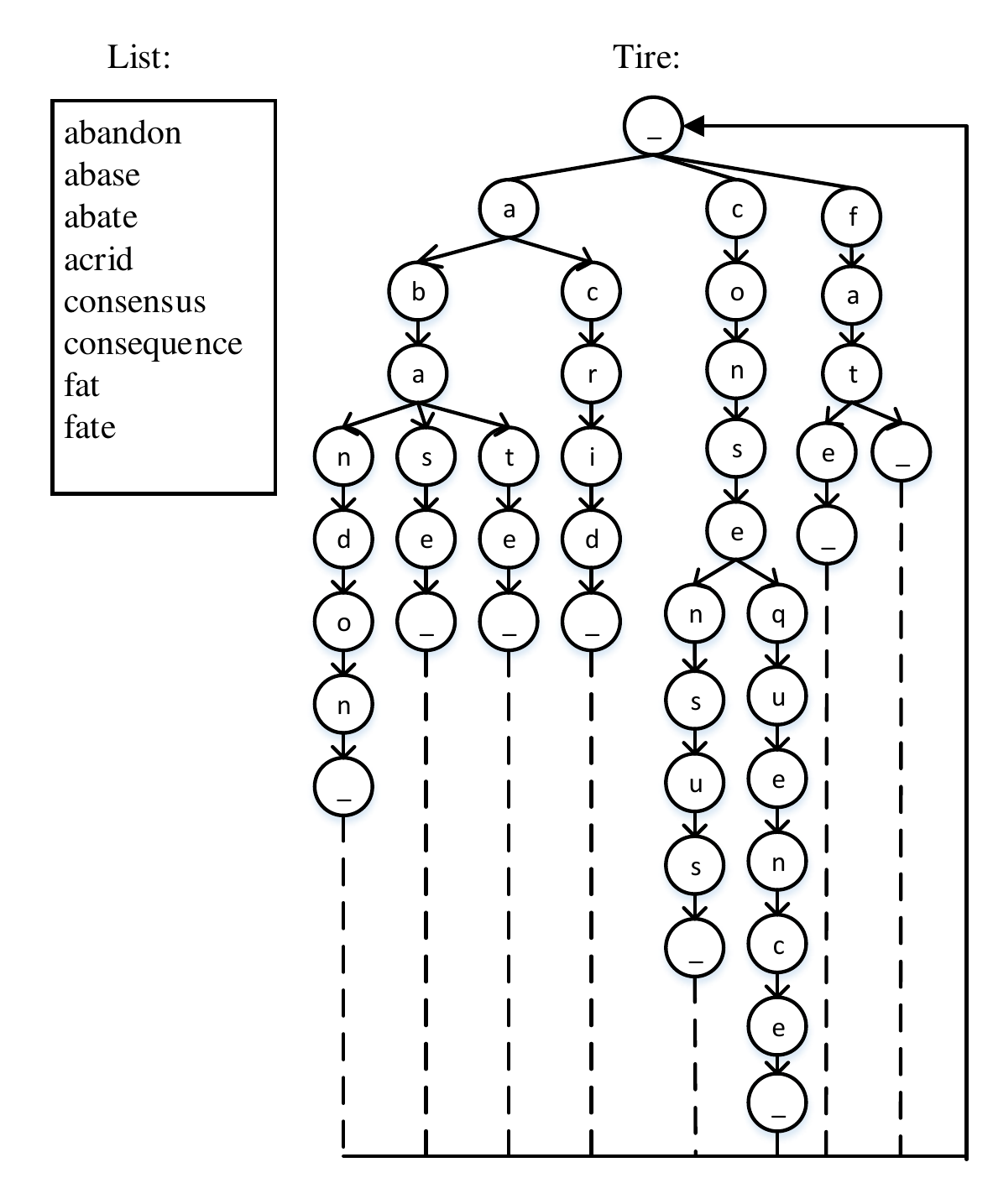}
  \caption{This dictionary includes 8 words. The trie is a tree. `\_' marks blank in English sentences.}
  \label{fig:trie}
\end{figure}

The straightforward way to store a dictionary is to list every word in it, with a lookup time complexity of $O(N\cdot S)$ ($N$ represents the number of words, and $S$ represents the length of the word). Apparently, by using this DS, Algorithm \ref{alg:all improvement} will perform poorly in the calculation of $Pr(k|\hat{B}(i))$ in the line 11. To reduce the time complexity, a trie can replace the list. An example is shown in Fig. \ref{fig:trie}.

The trie is a tree, and each node in the trie reflects a label. As an English dictionary, the label is just the character. Note that in this case $K$ is equal to $27$ because there are 26 characters in English plus the blank symbol. Make all the child nodes of each parent node in alphabetical order. The special label in the trie is the blank, separating every word in an English sentence. To distinguish it from `$\phi$', we use `\_' to mark it. Every word except the first one in an English sentence starts from a blank and ends with a blank, so each path in trie starting from the root node and ending with `\_' can describe a specific word. Notice that if a node $N_x$ has a child node which is `\_', the address of this child node is the same as the address of the root node. This means the `\_' at the end of each word does not actually take space. This mechanism enables us to search the tree circularly and save the memory space for the `\_' at the end of each word.

By shaping the dictionary into a trie, the time complexity of checking if $\boldsymbol{y}+k$ is in the dictionary is reduced. Define a dictionary pointer as $DP(i)$ of $\hat{B}(i)$ to mark the address of the last character in the last word in $\hat{B}(i)$. Every time when $Pr(k|\hat{B}(i))$ is calculated, we only need to find out if $k$ is one of the successors of the node which is pointed to by $DP(i)$. As a result, the lookup time complexity is reduced to $O(K)$. And each $DP(i)$ can be stored in $\hat{B}(i).SL$.

If the English dictionary with 191,735 words is stored as a trie, there will be 425,983 nodes in it. Each node may have at most $K$ ($K=27$) successors, and the address of each successor takes 19 bits ($\log_2 425983=18.7$).
To store a single node, the addresses of its all successors are in need. These addresses are stored , so that the storage structure can be designed as a matrix which has 425,983 rows and 27 columns. Because of the search direction in the trie (following the arrows in Fig. \ref{fig:trie}), the character which is represented by each node does not need to be saved in this matrix.
So the storage with an immediate way takes 425983$\times$27$\times$19=218,529,279bits=26.05MB.
The size of the trie is much smaller than that of n-gram LM, but it still can be futher compressed. Actually, the number of the successors of most of these nodes are smaller than $K$, so the matrix is a sparse one.
\subsection{Transform the Trie Into a Binary Tree}
The first compression strategy is reshaping the trie. A typical way to transform a multi-branched tree into a binary tree is to merely save a node's first child node and the first sibling from the right, so that each node will have only two child nodes.

A binary tree is well suited as the DS for the dictionary which is used in the CTC beam search decoding, because of the calculation order of $Pr(k|y)$ ($k=1,2,...,K$).

\begin{figure}[ht]
  \centering
  \includegraphics[scale=0.75]{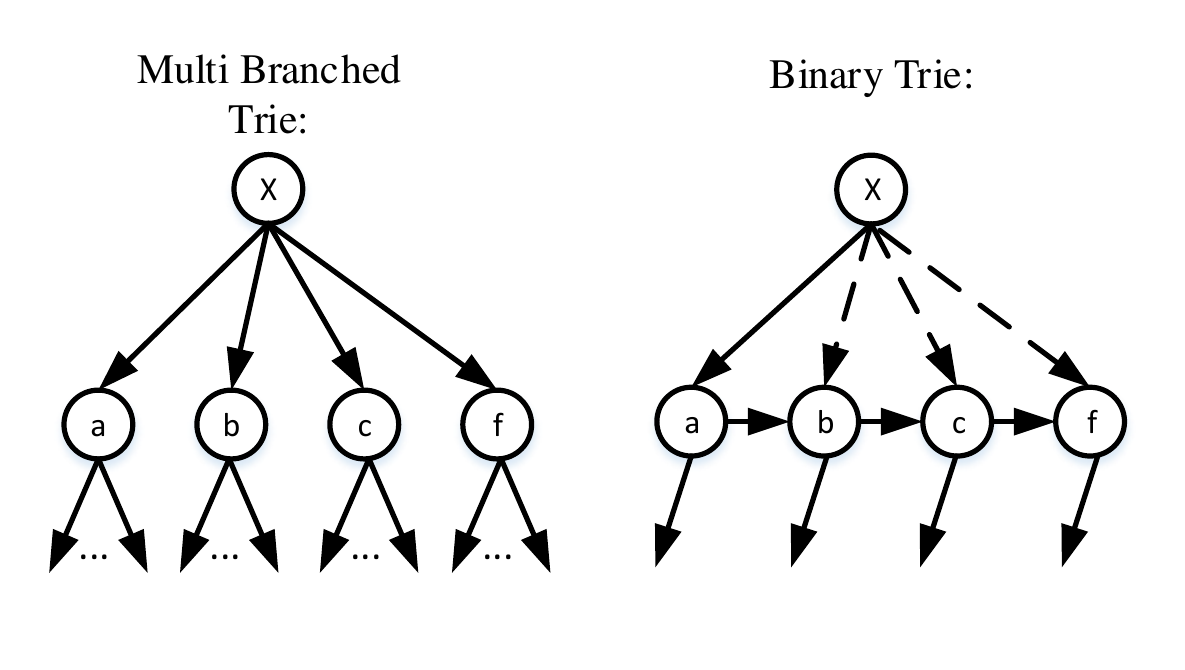}
  \caption{Transforming each node in the multi branched trie in this way will create a binary trie.}
  \label{fig:binary trie}
  \end{figure}

More importantly, a binary trie created by this means can save storage space. The transformation of the DS of trie is shown in Fig. \ref{fig:binary trie}. To store a node in the binary trie, the information in need includes which character this node represents where its left child (first child in the original trie) is and where its right child (first sibling from the right in the original trie) is. And they share the same address in memory space. As the character consumes 5 bits ($\log_227=4.75$), the storage space occupies 425983$\times$(5+2$\times$19)=18,317,269bits=2.18MB.

Another way to transform the multi-branched trie into a binary trie is to use the PATRICIA algorithm\cite{Morrison1988PATRICIA}. A Patricia tree is a special type of trie, highly-efficient in string matching. It is a more appropriate method for matching a single word, but not suitable for the decoding algorithm used in this paper.

\subsection{Compress the Address}
For each node, the addresses of its child nodes still occupy too much space. In this subsection, we compress the address of the left child first, and then we compress the other.

\subsubsection{Compress the Address of the Left Child}
All nodes in binary trie except the `\_' at the end of a word must have a left child, because every word ends with a blank. Assuming that each node is next to its first child in memory, a single bit is already enough to identify its left child: 1 represents that the left child is the `\_', and 0 represents that it is not the `\_'. To make sure every node is next to its left child, the preorder traversal of the binary trie should be stored in memory.

\subsubsection{Compress the Address of the Right Child}
The absolute address of the right child of node $N_x$ can be replaced with a relative address. The relative address is the difference between the address of the right child of $N_x$ and the absolute address of $N_x$. In the dictionary with 191,735 words, the maximum of this difference is 41,647. So the relative address takes 16 bits ($log_241647=15.35$).

\begin{figure}[ht]
\centering
\includegraphics[scale=0.7]{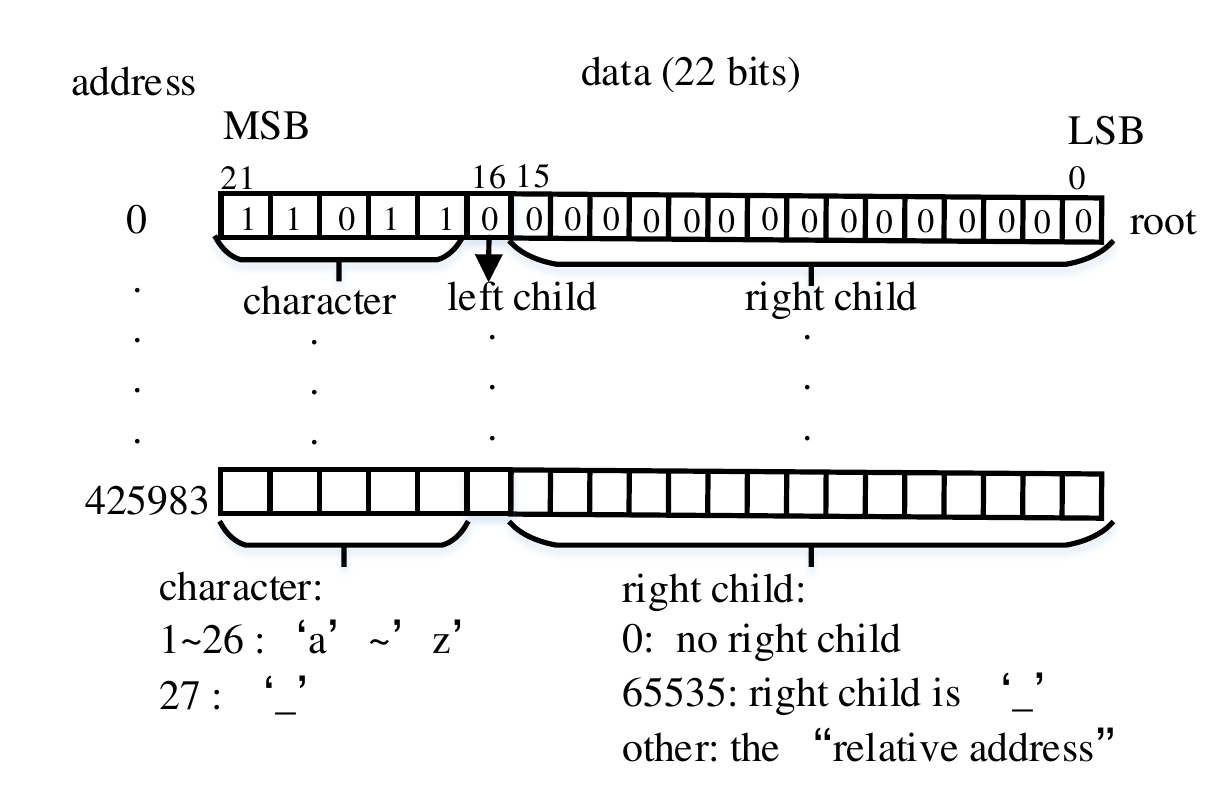}
\caption{The data storage format of the compressed dictionary. The root node is stored in address 0.}
\label{fig:dict storage}
\end{figure}

After compressing these addresses, the storage space decreases to 425983$\times$(5+1+16)=9,371,626bits=1.12MB.
The data storage format of the compressed dictionary is given in Fig. \ref{fig:dict storage}.

\subsection{Apply the Compressed Dictionary to Decoding}

\begin{figure}[ht]
\centering
\includegraphics[scale=0.6]{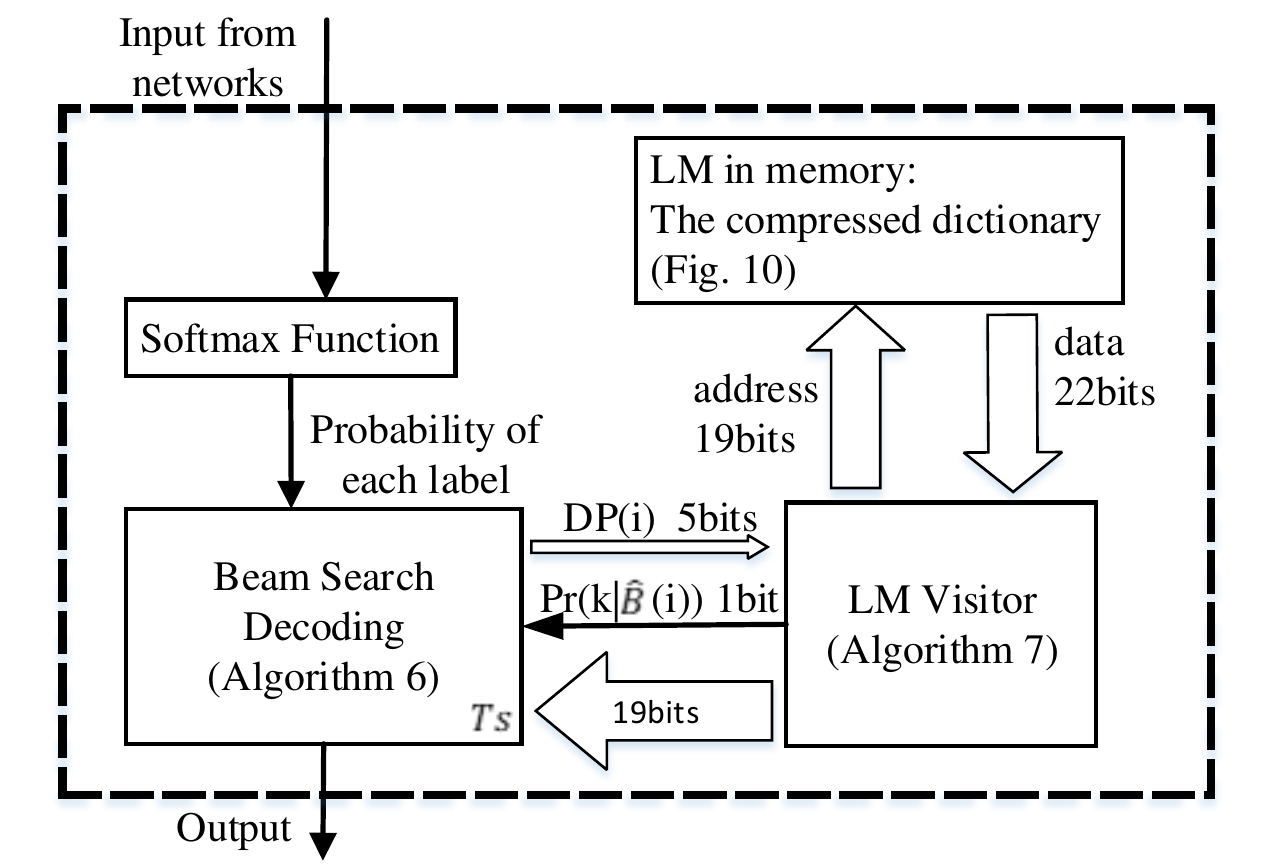}
\caption{The CTC-decoder designed by this work.}
\label{fig:final framework}
\end{figure}

To ensure the low dependency between the modules in the CTC-decoder, we set the LM Visitor module to control the access to LM. Algorithm \ref{alg:all improvement} leaves three interfaces to make connections with the LM Visitor, including $DP(i)$, $Pr(k|\hat{B}(i))$ and $T_S$. When the $Pr(k|\hat{B}(i))$ is calculated in the line 11 of Algorithm \ref{alg:all improvement}, the LM Visitor needs the value of $DP(i)$, and gives the value of $Pr(k|\hat{B}(i))$ back. Afterwards, the LM Visitor assigns the variable $T_S$ in the line 12 of Algorithm \ref{alg:all improvement}. Algorithm \ref{alg:LM} is used by the LM Visitor. The connections between Algorithm \ref{alg:all improvement}, Algorithm \ref{alg:LM} and various modules in the CTC-decoder are shown in Fig. \ref{fig:final framework}. Note that the constant $inv$ means the given address is invalid (at the same time, the $Pr(k,\hat{B}(i),t)$ must be zero).

\begin{algorithm}
\caption{LM Visitor}
\label{alg:LM}
\begin{algorithmic}[1]
\STATE const $inv=2^{19}-1=524287$
\STATE when a new $DP(i)$ reached :
\STATE $address\leftarrow DP(i)$
\STATE $flag\leftarrow0$
\STATE send $address$ to LM, get $data$ from LM
\IF {$data(16)=0$}
\STATE $address\leftarrow address+1$
\STATE send $address$ to LM, get $data$ from LM
\ELSE
\STATE $flag=2$
\ENDIF
\FOR {$k=1...26$}
\IF {$flag=0$ \AND $data(21:17)=k$}
\STATE $Pr(k|\hat{B}(i))\leftarrow1$
\STATE $T_S\leftarrow address$
\IF {$data(15:0)=0$}
\STATE $flag\leftarrow1$
\ELSE
\IF {$data(15:0)=65535$}
\STATE $flag\leftarrow2$
\ELSE
\STATE $address \leftarrow address+data(15:0)$
\STATE send $address$ to LM, get $data$ from LM
\ENDIF
\ENDIF
\ELSE
\STATE $Pr(k|\hat{B}(i))\leftarrow0$
\STATE $T_S\leftarrow inv$
\ENDIF
\ENDFOR
\STATE $k \leftarrow 27$
\IF {$flag=1$}
\STATE $Pr(k|\hat{B}(i))\leftarrow0$
\STATE $T_S\leftarrow inv$
\ELSE
\STATE $Pr(k|\hat{B}(i))\leftarrow1$
\STATE $T_S\leftarrow 0$
\ENDIF
\end{algorithmic}
\end{algorithm}

\begin{figure}[ht]
\centering
\includegraphics[scale=0.38]{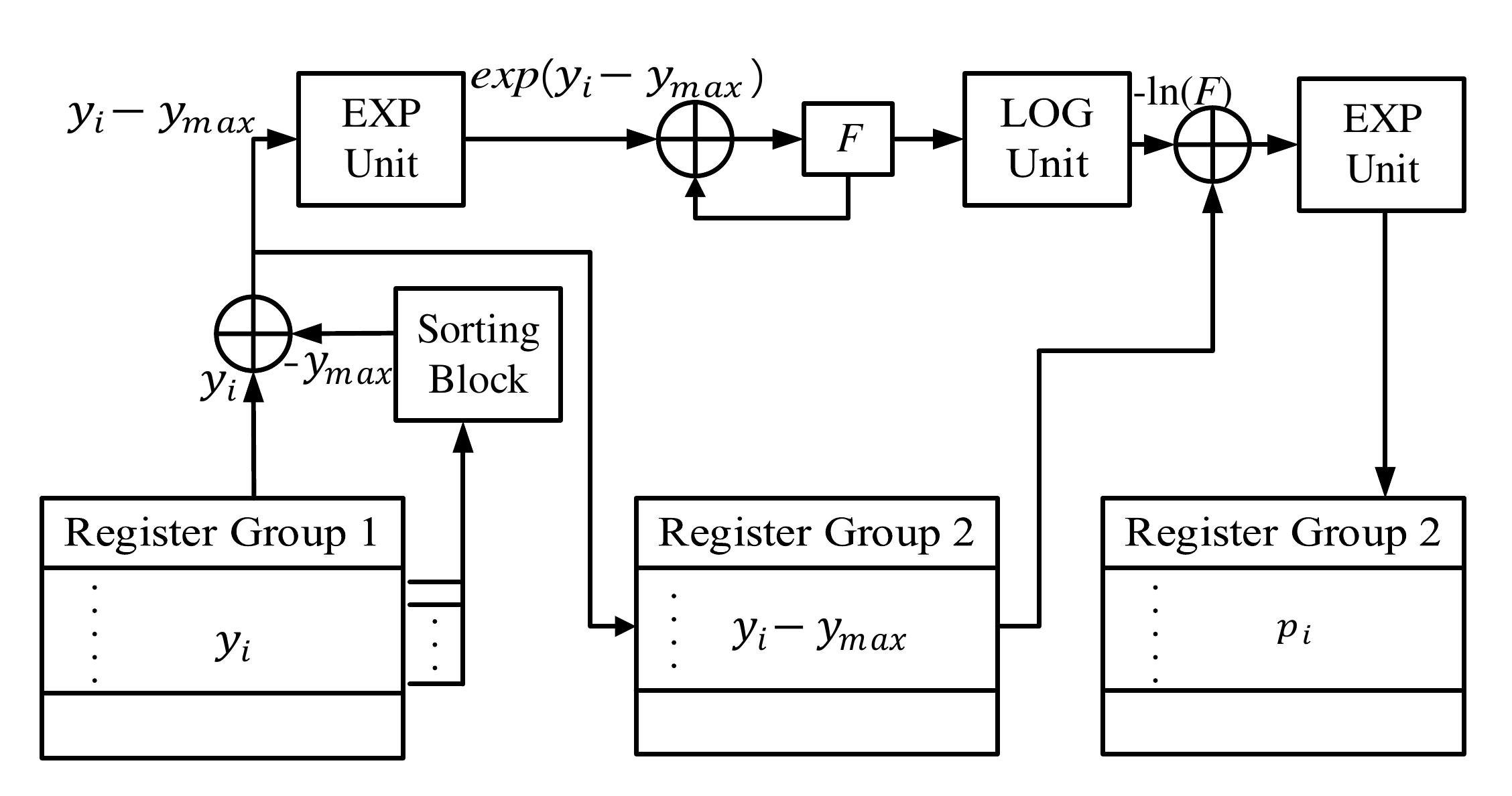}
\caption{The architecture of softmax in our CTC-deocder.}
\label{fig:softmax}
\end{figure}

\section{Experiment}
As mentioned earlier, we provide hardware-oriented and memory-efficient ways to implement every single module in the CTC-decoder shown in Fig. \ref{fig:final framework}.
The architecture of softmax functoin module  is shown in Fig. \ref{fig:softmax},
using several algorithmic strength reduction strategies described in Section \uppercase\expandafter{\romannumeral2}.
To demonstrate the advantages of our method, CTC-decoder is applied to a speech recognition task and a scene text recognition task. Meanwhile, we take a floating-point CTC-decoder using Algorithm \ref{alg:CTC Beam Search Decoding} as the baseline. Since there are some proper nouns and abbreviations in the transcriptions of these tasks, we add all words of datasets to our dictionary, and append $apostrophe$ in label list. The modification does not significantly affect original size and structure. 

\subsection{Speech Recognition Task}
In this experiment, we evaluate our method on a pre-trained Deep-speech-2 model\cite{Amodei2015Deep}, which is trained on LibriSpeech ASR corpus\cite{Panayotov2015Librispeech}. The WER is 11.27\% on ``test-clean'' set with greedy decoding used.\footnote{https://github.com/SeanNaren/deepspeech.pytorch/}
\subsubsection{Determine the Value of $W$}
\begin{figure}[ht]
\centering
\includegraphics[scale=0.20]{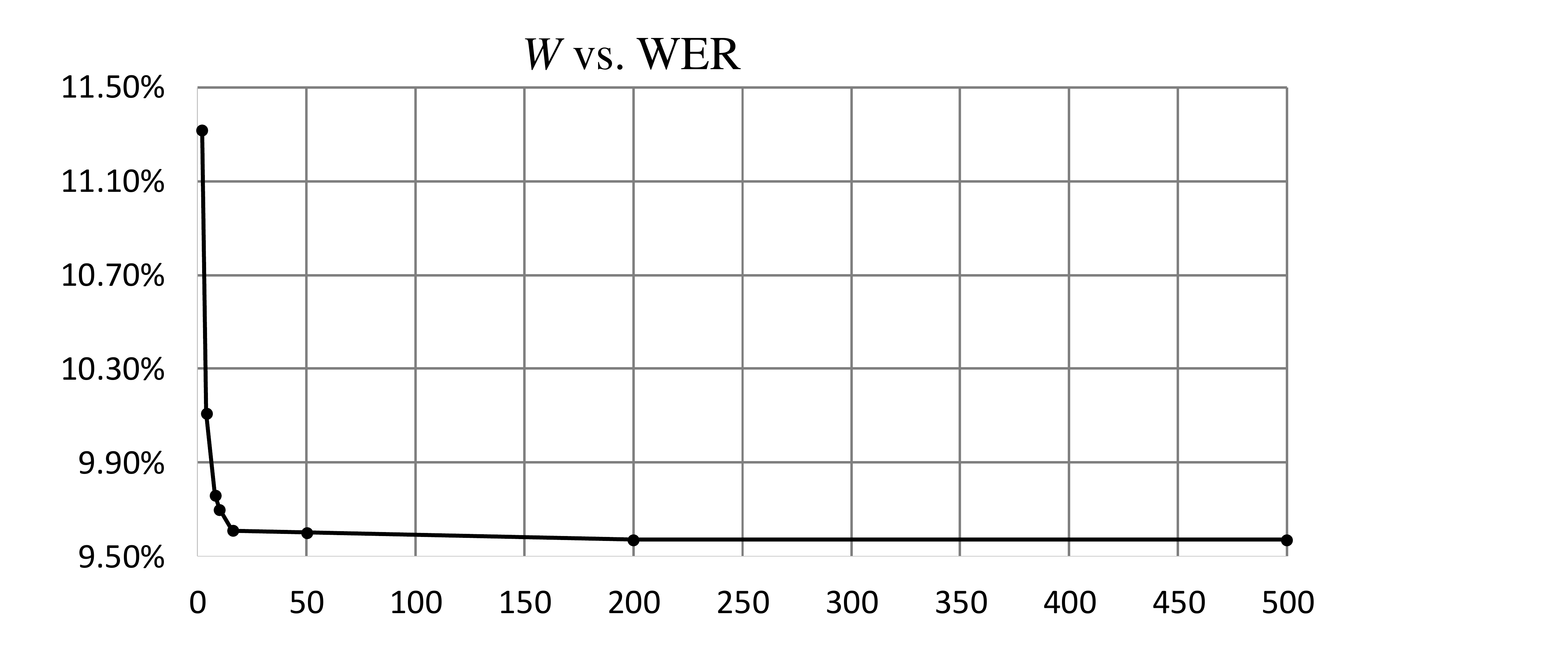}
\caption{The function relationship between $W$ and WER.}
\label{fig:W-WER}
\end{figure}

In Section \uppercase\expandafter{\romannumeral2}, the background of the beam search algorithm has been discussed. To balance the model size and performance, we conduct some experiments to evaluate the accuracy under different $W$. Fig. \ref{fig:final storage} illustrates the fact that the memory space used by Algorithm \ref{alg:all improvement} grows linearly as the data size increases.
The function of the size of $W$ vs. the word error rate (WER) of decoding is given in Fig. \ref{fig:W-WER}.

When $W<4$, the accuracy is unsatisfactory. When $W>50$, the calculation complexity becomes unacceptable while accuracy increases little. In addition, as the width of B1 is $\lceil \log_2W \rceil$, it is better that $w$ is an integral power of 2. At last, we choose 8 as the value of $W$.

\subsubsection{Fixed-Point Model of the Decoder}
After the determination of $W$, we build a model for a fixed-point CTC-decoder depicted in Fig. \ref{fig:final framework}.

\begin{figure}[ht]
\centering
\includegraphics[scale=0.20]{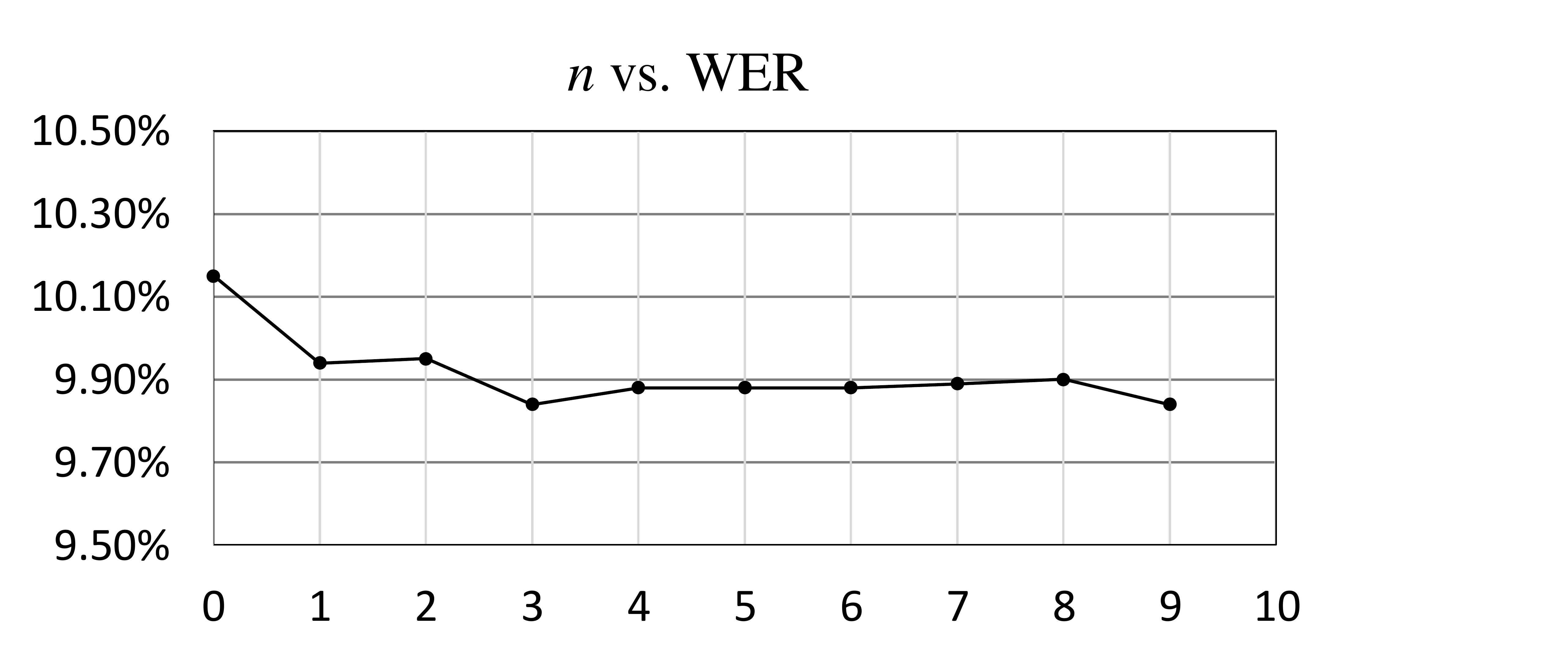}
\caption{The function relationship between $n$ and WER.}
\label{fig:n-WER}
\end{figure}

The number of integer bits is decided by the range of input $y_i$, while the number of fractional bits (denoted as $n$) is decided by the experiment. The value of $n$ has an impact on WER, and their functional relationship is depicted in Fig. \ref{fig:n-WER}. As a result, the input $y_i$ has eight bits: one sign bit, five integer bits and two fractional bits. 

In the experiment, we also find that the CTC-decoder does not require very accurate probabilities given by the softmax function. We can adjust the resolution of softmax function by three parameters: $\lambda$, $d_1$, $d_2$, where $\lambda$ is used to approximate the ratio of $e^x$ and $2^x$. We use $\lambda=1.5, 1/\lambda=0.625$ in this task, where 4-bits are used. The linear functions used in EXP Unit and LOG Unit may sacrifice the accuracy, but $d_1$ and $d_2$ can be adjusted to counteract this influence.
Some values of WER when $d_1$ and $d_2$ are set to different values are listed in TABLE \ref{table3}.

\begin{table}[H]
\normalsize
  \centering
  \caption{Some Values of WER When $d_1$ And $d_2$ Are Set to Different Values}
  \label{table3}

\begin{tabular}{ccc}
\hline
  $d_1$(binary) & $d_2$(binary) & WER\\
\hline
1.0000000110 & 0.1111111110 & 12.931\%\\
0.1111010001 & 0.1111111111 & 11.185\%\\
0.1101000001 & 0.1111111111 & 10.587\%\\
0.1011010110 & 0.1111110010 & 10.014\%\\
0.1011110111 & 0.1111110010 & 9.992\%\\
\hline
\end{tabular}
\end{table}

In addition, the probabilities calculated in the beam search decoding module also require fix-point processing. The experiment shows that if its decimal bit $q$ is less than 26, in some cases all the probabilities in $\hat{B}$ are smaller than $2^{-26}$, so they are all assigned to 0. To avoid this situation, we set $q$ equal to 30.

After the fix-point processing of softmax and the beam search decoding, a hardware-friendly model is created for softmax, replacing the most complex components by easy ones. With a greedy search strategy, we find a set of parameters for best WER, where $W=8$, $q=30$, $\lambda=1.5$, $1/\lambda=0.625$, $d_1=0.10111110111$, $d_2=0.1111110010$. The WER is 9.99\%, while 9.76\% in floating-point version. This minor loss of accuracy is generally acceptable.
Experiment results are shown in TABLE \ref{table4}. The baseline is a deepspeech2 model without a language model. $W=1$ means that CTC-Greedy Decoder is used. The floating-point and the fixed-point models share same configurations based on our method.
\begin{table}[H]
    \normalsize
    \centering
    \caption{Evaluation Results on LibriSpeech test-clean}
    \label{table4}

\begin{tabular}{ccc}
\hline
   Model & $W$ & WER \\
\hline
baseline(no LM) & 1 & 11.27\%\\
baseline(no LM) & 8 & 11.12 \% \\
floating-point & 8 & 9.76\%\\
fixed-point & 8 & 9.99\%\\
\hline
\end{tabular}
\end{table}

\subsection{Scene Text Recognition Task}
Synth90k dataset\cite{JaderbergSVZ14} is a synthetically generated dataset for text recognition, which consists of 9 million images covering 90k English words. We use a CRNN model\cite{Shi2017AnET} pre-trained on a subset of Synth90k dataset
\footnote{https://github.com/MaybeShewill-CV/CRNN\_Tensorflow}. A subset of dataset containing only character labels is used as test data.
Experiment results are shown in TABLE \ref{table5}, where the baseline is a CRNN model without a language model.

As mentioned above, we find the optimal quantization parameters in the same way. In this task, we choose parameter values with $\lambda=1/\lambda=1$, $d_1=0.1010111111$, $d_2=0.1111111111$, $q=30$, $W=8$. The final accuracy even increases from 90.85\% to 90.87\% when we convert the model from floating-point to fixed-point version.

\begin{table}[H]
\normalsize
  \centering
  \caption{Evaluation Results on Synth90k Dataset}
  \label{table5}
\begin{tabular}{ccc}
\hline
   Model & $W$ & Accuracy \\
\hline
baseline(no LM) & 1 & 47.47\%\\
baseline(no LM) & 8 & 88.02\%\\
floating-point & 8 & 90.85\%\\
fixed-point & 8 & 90.87\%\\
\hline
\end{tabular}
\end{table}

\subsection{Analysis of Applying Algorithm 6 to the Beam Search Decoding}
Section \uppercase\expandafter{\romannumeral4} improves the beam search decoding to reduce the memory space. In this subsection, we will figure out the compression ratio of the space used by the beam search decoding module in the English ASR task.

Each probability consumes 30 bits, and each SL consumes 19 bits (the address of a single node in the dictionary). Each Sentence has to store $T$ labels, while each label takes 5 bits ($\log_2K=\log_228=4.81$). According to Fig. \ref{fig:origin storage},  the original algorithm occupies $(109+5T)(KW+2W)=(26160+1200T)$ bits. According to Fig. \ref{fig:final storage}, Algorithm \ref{alg:all improvement} consumes $(2128+40T)$ bits. The results of each task are listed in TABLE \ref{table6}.

\begin{table}[H]
  \normalsize
  \centering
  \caption{Compression Ratio Results}
  \label{table6}
\begin{tabular}{ccc}
\hline
   Tasks & $T$ & Compression Ratio \\
\hline
ASR & 1800 & 29.49\\
STR & 25 & 17.95\\
\hline
\end{tabular}
\end{table}

Meanwhile, the experiments prove that the time of Algorithm \ref{alg:all improvement} spent in decoding (denoted as $\tau_2$) is less than the time spent by Algorithm \ref{alg:CTC Beam Search Decoding} (denoted as $\tau_1$). In our tests, when the number of the output vectors of softmax function is 697,310, we get $\tau_1=23.353$ seconds, and $\tau_2=20.816$ seconds.


\section{Conclusion and Future Work}
This paper has provided a hardware-oriented approach to build an CTC-decoder based on
an improved CTC beam search decoding.
The decoder has been implemented using C++ language and the experiments demonstrate that in English ASR tasks and STR tasks, the fixed-point CTC-decoder can save memory space for the beam decoding algorithm for 29.49 times and 17.95 times, respectively. The size of dictionary is compressed by 23 times. Additionally, there is little loss of precision compared with the floating-point CTC-decoder, and no increase is observed in computation time of the improved CTC beam search decoding.
In the future, a complete hardware implementation for the CTC-decoder will be conducted.

\appendices
\section{}

To reach the conclusion, we need to compare the probabilities adjusted by Algorithm \ref{alg:Adjust} with the original probabilities. To distinguish the adjusted probabilities from original ones, we use $\underline{Pr(\boldsymbol{y},t)}$,$\underline{Pr^+(\boldsymbol{y},t))}$ and $\underline{Pr^-(\boldsymbol{y},t))}$ to denote them.

Firstly, we assume that all the probabilities of $\hat{B}$ are enlarged by $\alpha_t$ at each time after update of $\hat{B}$.

Secondly, by using mathematical induction, Equation (\ref{eq:app_1}) can be proved.
\begin{equation}\label{eq:app_1}
    \begin{split}
    \forall t\in N^+,i\in \{1,2,...,W\},\exists!M_t>0:\\
    \begin{cases}
    \underline{Pr(\hat{B}(i),t)}=M_t\cdot Pr(\hat{B}(i),t),\\
    \underline{Pr^+(\hat{B}(i),t)}=M_t\cdot Pr^+(\hat{B}(i),t),\\
    \underline{Pr^-(\hat{B}(i),t)}=M_t\cdot Pr^-(\hat{B}(i),t).
    \end{cases}
    \end{split}
\end{equation}

The proof of Equation (\ref{eq:app_1}) can be expressed as follows:
\[
\begin{split}
\begin{aligned}
\text{(1)}& t=1, \forall i\in \{1,2,...,W\}:\\
&\begin{cases}
\underline{Pr(\hat{B}(i),1)}=\alpha_1\cdot Pr(\hat{B}(i),1),\\
\underline{Pr^+(\hat{B}(i),1)}=\alpha_1\cdot Pr^+(\hat{B}(i),1),\\
\underline{Pr^-(\hat{B}(i),1)}=\alpha_1\cdot Pr^-(\hat{B}(i),1).\\
\end{cases}\\
\text{(2)}& \text{Assume Equation (\ref{eq:app_1}) is true when $t=m$, so we have:}\\
&\begin{cases}
\underline{Pr(\hat{B}(i),m)}=M_m \cdot Pr(\hat{B}(i),m),\\
\underline{Pr^+(\hat{B}(i),m)}=M_m \cdot Pr^+(\hat{B}(i),m),\\
\underline{Pr^-(\hat{B}(i),m)}=M_m \cdot Pr^-(\hat{B}(i),m).\\
\end{cases}\\
&\text{Noticing line 17 and line 30 in Algorithm \ref{alg:all improvement}, the update of}\\
&\text{$Pr(\hat{B}(i),t)$ is based on the $W$ biggest probabilities from}\\
&\text{all $Temp$. Define $\underline{Temp},~\underline{Temp^+}$ and $\underline{Temp^-}$ as adjusted}\\
&\text{ones. Considering line 11 and line 29-31 in Algorithm}\\
&\text{6, they can be evaluated as:}\\
&~\underline{Temp^+}=M_m\cdot Temp^+,\underline{Temp^-}=M_m\cdot Temp^-\\
&~\underline{Temp}=\underline{Temp^+}+
\underline{Temp^-}=M_m\cdot Temp\\
&\text{As $M_m>0$, the judging results of  the inequalities in line}\\
&\text{17 and line 30 are the same with or without the adjustment.}\\
&\text{After the loop from line 28 to line 48, probabilities in $B$}\\
&\text{can be found as follows:}\\
&\begin{cases}
\underline{Pr(B(i),m+1)}=M_m \cdot Pr(B(i),m+1),\\
\underline{Pr^+(B(i),m+1)}=M_m \cdot Pr^+(B(i),m+1),\\
\underline{Pr^-(B(i),m+1)}=M_m \cdot Pr^-(B(i),m+1).\\
\end{cases}\\
&\text{So in line 51, when~} t=m+1, \forall i\in \{1,2,...,W\}:\\
&\begin{cases}
\underline{Pr(\hat{B}(i),m+1)}=M_m \cdot \alpha_{m+1}\cdot Pr(\hat{B}(i),m+1),\\
\underline{Pr^+(\hat{B}(i),m+1)}=M_m \cdot \alpha_{m+1}\cdot Pr^+(\hat{B}(i),m+1),\\
\underline{Pr^-(\hat{B}(i),m+1)}=M_m \cdot \alpha_{m+1}\cdot Pr^-(\hat{B}(i),m+1).\\
\end{cases}\\
& \text{Let~} M_{m+1}=M_m \cdot \alpha_{m+1},\\
&\begin{cases}
\underline{Pr(\hat{B}(i),m+1)}=M_{m+1}\cdot Pr(\hat{B}(i),m+1),\\
\underline{Pr^+(\hat{B}(i),m+1)}=M_{m+1}\cdot Pr^+(\hat{B}(i),m+1),\\
\underline{Pr^-(\hat{B}(i),m+1)}=M_{m+1}\cdot Pr^-(\hat{B}(i),m+1).
\end{cases}\\
& \text{So when $t=m+1$, Equation (\ref{eq:app_1}) is correct.}\\
\text{As} & \text{~a result, when $t\in N^+$, Equation (\ref{eq:app_1}) is correct.}
\end{aligned}
\end{split}
\]

Thirdly, by setting $t$ as $T$, the mathematical relationship between $Pr(\hat{B}(i),T)$ and $\underline{Pr(\hat{B}(i),T)}$ can be expressed as:
\begin{equation}\label{eq:app_2}
\begin{split}
\forall i\in \{1,2,...,W\},\exists M_T>0:\\
\underline{Pr(\hat{B}(i),T)}=M_T\cdot Pr(\hat{B}(i),T).
\end{split}
\end{equation}

Fourthly, set the maximum of $Pr(\hat{B}(i),T)$ as $Pr(\hat{B}(maxi),T)$:
\begin{equation}\label{eq:app_3}
\forall i\in \{1,2,...,W\}:Pr(\hat{B}(maxi),T)>Pr(\hat{B}(i),T).
\end{equation}
According to Equations (15) and (16), it can be shown that:
\begin{equation}\label{eq:app_4}
\begin{split}
\begin{aligned}
\forall i\in \{1,2,...,W\}:&\\
M_T\cdot Pr(\hat{B}(maxi),T)>&M_T\cdot Pr(\hat{B}(i),T),\\
\underline{Pr(\hat{B}(maxi),T)}>&\underline{Pr(\hat{B}(i),T)}.
\end{aligned}
\end{split}
\end{equation}
Finally, it is proved the maximum of $\underline{Pr(\hat{B}(i),T)}$ is still $\underline{Pr(\hat{B}(maxi),T)}$.

\bibliographystyle{plain}
\bibliography{bibitem}


\end{document}